\documentclass[12pt]{article}
\usepackage{amsmath,bm,mathtools,amsthm}
\usepackage{graphicx,psfrag,epsf}
\usepackage{enumerate}
\usepackage{natbib}
\usepackage{url} 
\usepackage{soul,xcolor}
\usepackage{setspace}

\setstcolor{red}

\DeclareMathAlphabet\mathbfcal{OMS}{cmsy}{b}{n}

\newcommand{\blind}{0}

\begin{document}

\def\spacingset#1{\renewcommand{\baselinestretch}%
{#1}\small\normalsize} \spacingset{1}

\numberwithin{equation}{section}
\newtheorem{thm}{Theorem}
\newtheorem{proposition}{Proposition}
\newtheorem{cor}{Corollary}[thm]
\newtheorem*{rmk}{Remark}
\newtheorem{result}{Result}
\newtheorem{definition}{Definition}
\newtheorem*{definition*}{Definition}
\newcommand{\ignore}[1]{}


\if0\blind
{
  \title{\bf A Study on the Power Parameter in Power Prior Bayesian Analysis
  }
  \author{Zifei Han\hspace{.2cm}\\
    University of International Business and Economics\\
    zifeihan@uibe.edu.cn\\
     \\
    Keying Ye \\
        The University of Texas at San Antonio \\
     keying.ye@utsa.edu\\
     \\
    Min Wang\footnote{Corresponding author: min.wang3@utsa.edu} \\
    The University of Texas at San Antonio\\
    min.wang3@utsa.edu}
  \maketitle
} \fi

\if1\blind
{
  \bigskip
  \bigskip
  \bigskip
  \begin{center}
    {\LARGE\bf Title}
\end{center}
  \medskip
} \fi

\bigskip
\begin{abstract}
The {\it power prior} and its variations have been proven to be a useful class of informative priors in Bayesian inference due to their flexibility in incorporating the historical information by raising the likelihood of the historical data 
to a fractional power $\delta$. 
The derivation of the marginal likelihood  based on the original power prior, and its variation, the normalized power prior,  introduces 
a scaling factor $C(\delta)$ in the form of a prior predictive distribution with powered likelihood. In this paper, we show that the scaling factor might be infinite for some positive $\delta$ with conventionally used initial priors, which would change the admissible set of the power parameter. This result seems to have been almost completely 
ignored in the literature. We then 
illustrate that such a phenomenon may 
jeopardize  the posterior inference under the power priors when the 
initial prior of the model parameters is improper. The main findings of this paper suggest that special attention should be paid when the suggested level of borrowing is close to $0$, while the actual optimum might be below 
the suggested value. 
We use a normal linear model as an example for illustrative purposes.  
\end{abstract}

\noindent%
\vfill

\newpage
\spacingset{1.45} 
\section{Introduction}
\label{sec:intro}

The power prior is a class of informative priors constructed from historical data in Bayesian inference. It allows researchers to incorporate historical information via the likelihood function of the historical data raised to a power. 
The basic principle is to use a {\it power parameter} $\delta$ ($0 \leq \delta \leq 1$) 
to control the influence of the historical data on the current study.
This information borrowing strategy was introduced by 
\cite{IbrahimChen98, IbrahimChen00} with the formulation 
\begin{equation}
\label{eq:pp}
\pi(\bm{\theta} \mid D_{0},\delta) \propto
L(\bm{\theta} \mid D_{0})^{\delta} \pi_{0}(\bm{\theta}), 
\end{equation}
where $L(\bm{\theta} | D_{0})$ is the likelihood based on the historical data $D_{0}$, $\pi_{0}(\bm{\theta})$ is an {\it initial prior} for the parameter of interest $\bm{\theta}$, and $\delta$ may be assumed to be fixed. 
Despite the fact that $\pi({\bm \theta} | D_{0}, \delta)$ depends   
on $D_{0}$ and $\delta$, 
it is often treated as an informative {\it prior} for the current study and is essentially a 
powered posterior based on $D_0$. It is worth noting that the power prior in (\ref{eq:pp}) and the initial prior $\pi_{0}({\bm \theta})$ are not necessarily proper, 
so long as the resulting posterior is proper. In many practical applications, it seems natural to expect most information to come  from the historical data rather than the initial prior, and thus, we frequently adopt non--informative priors such as reference priors \citep{Berg:Bern:Sun:2009} or weakly informative priors as the initial prior in Bayesian analyses.
Assuming that the likelihood based on the current data $D$ is of the form $L(\bm{\theta} | D)$,
using the power {\it prior} defined in (\ref{eq:pp}), the {\it posterior} of ${\bm \theta}$ when $\delta$ is fixed has the form 
\begin{equation}
\pi({\bm \theta} \mid D, D_0, \delta) \propto L(\bm{\theta} \mid D) \pi(\bm{\theta} \mid D_{0},\delta),
\label{eq:pp-posterior}
\end{equation}
where $\pi(\bm{\theta} | D_{0},\delta)$ is considered as a prior based on the historical data.

An important issue in the use of the power prior is to 
determine the level of borrowing  by selecting a 
sensible power parameter, which can usually be determined either 
by expert opinions in advance, or using criteria that can reflect the prior--data conflict. \cite{Ibrahim15} proposed multiple information--based criteria to determine the level of borrowing based on the data compatibility. 
Among those criteria, 
the marginal likelihood criterion corresponding 
to the empirical Bayes solution has recently been extensively studied in the literature
\citep{Gravestock17,Nikolakopoulos18,Gravestock19,
Wiesenfarth20,Ollier20,Pateras21}. 
To derive the marginal likelihood, 
one first integrates out the model parameter 
${\bm \theta}$ from the right hand side of  (\ref{eq:pp-posterior}) to obtain 
$$\int_{\bm{\Theta}} L(\bm{\theta} \mid D) 
\pi(\bm{\theta} \mid D_{0},\delta) d \bm{\theta} 
= 
\frac{\int_{\bm \Theta} L(\bm{\theta} \mid D) L(\bm{\theta} \mid D_0)^{\delta} 
\pi_{0}(\bm{\theta})  d\bm{\theta}}{
C(\delta)},
$$
where $\pi(\bm{\theta} | D_{0},\delta)$
is the power prior of the form (\ref{eq:pp}), with a normalizing constant 
$C(\delta) = 
\int_{\bm{\Theta}}
L(\bm{\theta} | D_{0})^\delta
\pi_{0}(\bm{\theta}) d\bm{\theta}$, and 
$\bm{\Theta}$ is the parameter space of $\bm{\theta}$. 
Since $C(\delta)$ is a function of $\delta$, for the purpose of selecting a power parameter, one cannot drop it when comparing different models indexed by $\delta$. 
To avoid an infinite $C(\delta)$, 
we denote the marginal likelihood as 
\begin{equation}
\label{eq:marginal-lik}
m(\delta \mid   D_{0}, D) = 
\frac{\int_{\bm \Theta} L(\bm{\theta} \mid D) L(\bm{\theta} \mid D_0)^{\delta} 
\pi_{0}(\bm{\theta})  d\bm{\theta}}{
C(\delta)}I_{\mathcal{A}}(\delta),  
\end{equation}
in which the indicator function 
$I_{\mathcal{A}}(\delta) = 1$ if $\delta \in \mathcal{A}$ and $0$ otherwise, and the non--empty set is given by 
\begin{equation}
\label{eq:A-define}
\mathcal{A} = \{\delta  \mid   
\delta \in [0, 1] ~{\rm and} ~ C(\delta) < \infty\}.\end{equation}
 We call $\mathcal{A}$ the {\it feasible set} of the power parameter $\delta$, which is  {\it complete} when $\mathbfcal{A} = [0,1]$, and {\it semi--complete} if $\mathbfcal{A} = (0,1]$ and otherwise {\it incomplete}. Then the suggested level of borrowing using the empirical Bayes solution could be written as 
$\delta_{\rm EB} = \arg\max_{\delta \in \mathcal{A}} m(\delta | D_{0}, D)$ \citep{Gravestock17}, in which the condition of $\delta \in \mathcal{A}$ to ensure a finite $C(\delta)$ in (\ref{eq:A-define}) seems to be ignored. Consequently,  
this might influence the marginal likelihood in (\ref{eq:marginal-lik}) for quantifying the data compatibility. 
Although the formulation guarantees that $\delta_{\rm EB} \in \mathcal{A}$, the domain of  $\mathcal{A}$ might be  incomplete, and can thus preclude an optimal level of borrowing. 


On the other hand, as a natural Bayesian response to the uncertainty of  $\delta$, we can assign a hyperprior $\pi_0(\delta)$, which results in a hierarchical
power prior. 
When a weakly--informative prior $\pi_0(\delta)$  is used, hypothetically, the
posterior of $\delta$ would reflect the data compatibility in a 
semi--automatic way. Therefore \cite{IbrahimChen00} proposed the {\it joint power prior} by 
specifying a prior for $({\bm \theta}, \delta)$ jointly with the form
\begin{equation}
\label{eq:jpp}
\pi(\bm{\theta}, \delta \mid  D_{0}) \propto
L(\bm{\theta} \mid D_{0})^{\delta}
\pi_{0}(\bm{\theta}) \pi_{0}(\delta). 
\end{equation}
It is noted that with the prior in (\ref{eq:jpp}), the likelihood 
principle \citep{Birnbaum62} is violated, since using various forms of
the historical likelihood differ by a multiplicative positive constant $c_0$ would result in different posteriors (Their posteriors would be differed by 
$c_0^{\delta}$; see discussions in \citealp{Neuenschwander09} and \citealp{Ye22}). 
We provide a simple example in the Appendix to illustrate its consequence. 
The prior in (\ref{eq:jpp}) simply specifies a joint prior of $({\bm \theta}, \delta)$ directly  \citep{Ibrahim15}. 
 If one first specifies a prior $\pi_0(\delta)$,
 then specifies a conditional prior of 
 ${\bm \theta}$ given $\delta$ using the power prior in (\ref{eq:pp}), the normalizing factor  $C(\delta)$ in $\pi({\bm \theta} | D_0, \delta)$
 should not be dropped as well. This is due to the fact that the power $\delta$ is treated as a parameter. Therefore \cite{Duan06a} proposed the following modified power prior, named as the {\it normalized power prior}, given by 
\begin{equation}
\label{eq:npp-redef}
\pi(\bm{\theta},\delta \mid D_{0}) \propto 
\frac{L(\bm{\theta} \mid D_{0})^\delta
\pi_{0}(\bm{\theta})\pi_{0}(\delta)
I_{\mathcal{A}}(\delta)}{C(\delta)},
\end{equation}
which obviously obeys the likelihood principle since a multiplicative constant before $L(\bm{\theta} | D_{0})$ will be canceled out \citep{Neuenschwander09}. This prior exists under a non--restrictive assumption of a non--empty set for $\mathcal{A}$, since $1 \in \mathcal{A}$ 
as long as the initial prior 
leads to a proper posterior in a conventional Bayesian analysis. 
Of particular note is that in the current literature, almost all normalized power prior formulae omit  $I_{\mathcal{A}}(\delta)$ and assume  either $\delta \in [0, 1]$ with a proper prior $\pi_{0}({\bm \theta})$ or $\delta \in (0, 1]$ with an improper prior $\pi_{0}({\bm \theta})$. In this study, we will show that 
the latter may not be the case 
even for the commonly used  
improper priors that can yield proper posteriors, 
such as some reference priors. This finding justifies the importance of 
studying the range of the admissible $\delta$ and/or the feasible set $\mathbfcal{A}$ in the marginal likelihood (\ref{eq:marginal-lik}) and the normalized power prior (\ref{eq:npp-redef}). However, such importance has not 
been given sufficient consideration in the past. 

The major contribution of this study is to 
 mathematically examine an important but almost completely ignored key point that, 
with a commonly used improper initial prior that is believed to be objective, the feasible parameter space $\mathcal{A}$ of $\delta$ might be restricted to only a 
subregion of $(0,1]$, i.e., $\mathbfcal{A}$ may be incomplete. 
With the exception of \cite{Duan06a}, almost all the research articles by default assume $\delta \in (0, 1]$. 
As a result, either the marginal likelihood in (\ref{eq:marginal-lik}) or the  normalized power prior (\ref{eq:npp-redef}) might not be able to accurately quantify the data compatibility. 
This is because $\mathcal{A}$ could possibly 
exclude the optimal range of $\delta$. 
We further prove that, under certain conditions, the feasible set $\mathcal{A}$ is a convex set with 
a lower limit  $\delta^{*} \geq 0$, which suggests 
that the phenomenon described above might happen 
when the heterogeneity between $D_0$ and $D$ is strong. For instance, when the optimal level of borrowing under strong heterogeneity is close to 0, the set 
$\mathbfcal{A}$ can exclude it by definition. 
Albeit its impact on the analysis could be mild in many scenarios, when using the empirical Bayes or the normalized power prior, 
researchers should be vigilant against the use of certain predominantly used improper priors as the initial objective prior, such as the reference prior in a normal linear model.  The root cause is that, 
despite the improper prior with a full likelihood can yield a 
proper posterior, the same prior with a fractional likelihood may 
lead to an improper posterior for some $\delta$.

The rest of the paper is organized as follows. In Section \ref{sec:meth}, 
we establish theoretical results regarding the propriety of the power priors 
in general. Explicit 
results under the normal linear model are provided in 
Section \ref{sec:linear-model} with several commonly used initial priors. 
In Section \ref{sec:simulation}, 
we conduct a numerical study to illustrate the 
undesirable behavior when using the power prior with empirical Bayes 
or the normalized power prior in a normal linear model. 
We further discuss the implications of our results 
in Section \ref{sec:conc}.

\section{Propriety of the Power Priors}
\label{sec:meth}
In this section, we discuss some fundamental issues regarding the 
propriety of the power priors in general, which shed light on the 
form of a feasible set $\mathcal{A}$ defined in (\ref{eq:A-define}). 
Proofs of the theorems are provided in the Appendix. 
For all of the following results, we consider the  power prior model with the form 
$\pi(\bm{\theta} | D_{0},\delta) \propto
L(\bm{\theta} | D_{0})^{\delta} \pi_{0}(\bm{\theta})$ as described in 
(\ref{eq:pp}). 

\begin{thm}\label{thm:proper-prior}
When the initial prior $\pi_{0}({\bm \theta})$ is proper, the power prior  
is always proper with a 
non--negative $\delta$. 
Therefore the range of feasible $\delta$ defined 
in (\ref{eq:marginal-lik}) is complete, i.e., $\mathcal{A} = [0, 1]$. 
\end{thm}

It is well known that with a proper prior and a non--degenerate (full) likelihood, 
the set of observations in which the posterior is improper is 
a Lebesgue null set. This guarantees that the propriety of the posterior holds almost surely 
with a proper prior. Theorem \ref{thm:proper-prior} 
demonstrates that 
using a proper (initial) prior $\pi_{0}({\bm \theta})$ 
with a fractional likelihood would be a safe choice since it can also 
almost surely guarantee posterior propriety. 
However, we show in the following theorem that, 
this may not be the case when an improper (initial) prior of $\boldsymbol{\theta}$ is used.

\begin{rmk} \label{rmk:improper-prior}
When the initial prior $\pi_{0}({\bm \theta})$ is improper, 
the power prior may or may not be proper, 
even if the posterior with the corresponding full likelihood is proper. 
In other words, $\int_{\bm{\Theta}} \pi_{0}({\bm \theta}) 
L({\bm \theta}  |  D_0) d\bm{\theta}< \infty$
does not necessarily indicate 
$\int_{\bm{\Theta}} \pi_{0}({\bm \theta}) 
L({\bm \theta}  |  D_0)^{\delta} d\bm{\theta} < \infty$ for all $\delta > 0$. 
\end{rmk}

Note that unlike the proof of Theorem \ref{thm:proper-prior},
here Jensen's inequality can no longer guarantee a finite 
upper bound for an improper density $\pi_{0}({\bm \theta})$. 
More details can be found in the Appendix.
A primary example is also provided in Section \ref{sec:linear-model} for the 
 normal linear model. 
This suggests that when using the marginal likelihood criterion (\ref{eq:marginal-lik}) or the 
normalized power prior (\ref{eq:npp-redef})
with an improper initial prior on ${\bm \theta}$, 
the feasible set $\mathcal{A}$ might be 
restricted to a subregion of $(0, 1]$, which is incomplete. 
To better understand their properties, 
we have the following theorem. 

\begin{thm}\label{rmk:improper-prior-proper-region}
If  
the power prior with a positive power parameter 
$\delta^{*}$ is proper, 
then for all $\delta \geq \delta^{*}$, 
$\pi_{0}({\bm \theta}) L({\bm \theta}  |  D_0)^{\delta}$ is integrable.
\end{thm}

Based on the Theorem \ref{rmk:improper-prior-proper-region} 
and all the related results above, we come up with the following 
corollary regarding the feasible parameter space $\mathcal{A}$ of $\delta$ in the formulation of the  
marginal likelihood (\ref{eq:marginal-lik}) 
and the normalized power prior (\ref{eq:npp-redef}). 

\begin{cor}
\label{thm:par-space-npp}
The feasible parameter space $\mathcal{A}$ of $\delta$ in the 
marginal likelihood (\ref{eq:marginal-lik}) and the 
normalized power prior (\ref{eq:npp-redef}) is 
a convex set with upper limit $1$ and 
lower limit $\delta^{*}$, where $0 \leq \delta^{*} \leq 1$.

\end{cor}

\section{Investigation on the Feasible Set for the Normal Linear Model}
\label{sec:linear-model}
In this section we provide some results under the commonly used 
normal linear model, with detailed derivations in the Appendix. 
These results will serve as 
examples to illustrate some of the findings obtained in Section 
\ref{sec:meth}, and will be further used for a numerical 
study in Section \ref{sec:simulation}. 
We derive the results for normal linear regression model with a common
unknown variance $\sigma^2$, while the result for a known variance is given by \cite{Ibrahim15}. The model is specified as 
 \begin{eqnarray*}
\bm{Y}=\mathbf{X}\bm{\beta} +\bm{\epsilon},\text{ with }\bm{\epsilon}\sim {\rm N}_{n}(\bm{0}_{n},\sigma^2 I_n),
\end{eqnarray*} 
where $I_n$ is the $n\times n$ identity matrix, ${\rm N}_{n}(\bm{0}_{n},\sigma^2 I_n)$ denotes the $n$--dimensional multivariate normal distribution with 
mean $\bm{0}_{n}$, the $n$--dimensional column vector of zeroes, and 
covariance matrix $\sigma^2 I_n$. 
The dimension of the vector $\bm{Y}$ is also $n$ and that of $\bm{\beta}$ is $p$. Similarly, for historical data 
we assume $\bm{Y}_0 = \mathbf{X}_0\bm{\beta} +\bm{\epsilon}_0$, with $\bm{\epsilon}_0\sim 
{\rm N}_{n_0}(\bm{0}_{n_0},\sigma^2 I_{n_0})$, and  both $\mathbf{X}_0'\mathbf{X}_0$ and $\mathbf{X}'\mathbf{X}$ are positive definite. 
We adopt similar notations to 
\cite{Ye22} but consider the prior with a more general form 
\begin{equation}
\label{eq:prior-general}
\pi_{0}({\bm \beta}, \sigma^2) \propto 
\frac{1}{(\sigma^2)^{t}}
\exp\left\{-\frac{1}{\sigma^2} 
\left[b + \frac{k}{2} ({\bm \beta} - {\bm \mu}_0)' {\bm R} 
({\bm \beta} - {\bm \mu}_0) \right]
\right\},
\end{equation}
where ${\bm R}$ is a known $p \times p$ real--valued 
positive--definite matrix, 
${\bm \mu}_0$ is a known $p \times 1$ real vector, 
$t$ and $b$ are non--negative real numbers, 
$k = 0$ or $1$. 
This class of prior includes several 
commonly used ones in the linear model as special cases. 
For example, when 
$t > 1+p/2, b > 0, k = 1$ and $\bm{R}$ is positive definite, 
the prior (\ref{eq:prior-general}) is the proper conjugate multivariate 
normal-inverse-gamma prior denoted as  
${\rm  N}_{p} \mbox{-} \Gamma^{-1} ({\bm \mu}_{0}, 
{\bm R}, t-p/2-1, b)$. If $t = 1+p/2, b = 0, 
k = 1$, and matrix $\bm{R} = g^{-1}(\mathbf{X}'\mathbf{X})$ with  
$g$ is a positive constant, 
the prior has the form 
$$\pi_{0}({\bm \beta}, \sigma^2) \propto 
\left(\frac{1}{\sigma^2}\right)^{\frac{p}{2}+1}
\exp\left\{-\frac{({\bm \beta} - {\bm \mu}_0)' {\mathbf{X}'\mathbf{X}}({\bm \beta} - {\bm \mu}_0) }{2g\sigma^2}\right\},$$
which is the well--known Zellner's $g$--prior 
\citep{Zellner:1986}. 
When $t = 1, k = b = 0$, the general form reduces to a reference prior $\pi_0({\bm \beta}, \sigma^2) \propto 1/\sigma^2$ \citep{Berg:Bern:Sun:2009}.

\begin{result}
\label{thm:normal-model}
Consider the initial prior of the form (\ref{eq:prior-general}) 
and assume $n_0 > p$. Then the power prior is proper when $\delta >  (2-2t+p)/n_0$.
 Specifically, 
\begin{enumerate}
\item[(a)] If we use the reference prior indicated above, 
then the power prior is proper only when $\delta \in \left(p/n_{0}, 1 \right]$.

\item[(b)] 
If the initial prior (\ref{eq:prior-general}) satisfies 
$t \geq 1+p/2$, then the power prior is proper for 
all $\delta \in (0, 1]$. 
This includes Zellner's $g$--prior and the 
proper conjugate normal-inverse-gamma prior.
\end{enumerate}
\end{result}

Next, we provide some closed--form results to guide  
the choice of $\delta$ when we use the power prior with a deterministic 
information--based criterion. In addition to the empirical Bayes approach 
described above based on the marginal likelihood criterion in (\ref{eq:marginal-lik}), we include another information criterion, 
the deviance information criterion (DIC) \citep{Spiegelhalter02}, 
which is extensively used for  model selection in 
Bayesian statistics. For notational simplicity, we define 
\begin{align*}
\hat{\bm{\beta}}_0& = (\mathbf{X}_{0}'\mathbf{X}_{0})^{-1}
\mathbf{X}_{0}'\bm{Y}_{0}, ~~
{S}_0 = (\bm{Y}_0-\mathbf{X}_{0} \hat{\bm{\beta}}_0)'(\bm{Y}_0-\mathbf{X}_{0} \hat{\bm{\beta}}_0),\\
\hat{\bm{\beta}}& = 
(\mathbf{X'X})^{-1}\mathbf{X}'\bm{Y}, \text{  and }
{S} = (\bm{Y}-\mathbf{X} \hat{\bm{\beta}})'(\bm{Y}-\mathbf{X} \hat{\bm{\beta}}).
\end{align*} 
Then we have the following result. 

\begin{result}
\label{thm:normal-mlcdic}
Consider the initial prior 
$\pi_{0}({\bm \beta}, \sigma^2)$ 
of the form in (\ref{eq:prior-general}) and assume $n_0 > p$. 
\begin{enumerate}
\item[(a)] 
The marginal likelihood in (\ref{eq:marginal-lik}) is of 
the form 
\begin{equation}
m(\delta \mid  D_{0}, D) \propto 
\frac{\Gamma(\nu) |\delta \mathbf{X}_{0}'\mathbf{X}_{0}
+ k {\bm R}|^{\frac{1}{2}} H_{0}(\delta)^{\nu_0}}
{\Gamma(\nu_0) |\mathbf{X}'\mathbf{X} + \delta 
\mathbf{X}_{0}'\mathbf{X}_{0}+ k {\bm R}|^{\frac{1}{2}}
H(\delta)^{\nu}}. 
\label{eq:marginal-lik-LM}
\end{equation}

\item[(b)] 
The DIC for a model with a certain value of $\delta$, denoted as 
${\rm DIC}(\delta|D_{0}, D)$, up to a constant, 
can be expressed as
\begin{align}
{\rm DIC}(\delta \mid  D_{0}, D) &=  
n\left\{ \log\left( \nu - 1\right) + 
\log H(\delta) 
- 2\psi(\nu) \right\}
+ \frac{\nu+1}{H(\delta)}
\left\{(\bm{\beta}^{*} -\hat{\bm{\beta}})'
\mathbf{X}'\mathbf{X}(\bm{\beta}^{*} -\hat{\bm{\beta}})+S\right\}
\nonumber \\
& ~~~ +
2{\rm tr}(\mathbf{X}'\mathbf{X} (\mathbf{X}'\mathbf{X} + \delta \mathbf{X}_0'\mathbf{X}_{0} +k \bm{R})^{-1}).
\label{eq:DIC-LM}
\end{align}
\end{enumerate}
In both parts,  
$\nu_0 = \frac{n_0 \delta -p}{2}+t-1$, 
$\nu = \nu_0 + \frac{n}{2}$, 
$\tilde{\bm{\beta}}=\left(\delta\mathbf{X}_0'\mathbf{X}_0 + k\bm{R} \right)^{-1}
\left(\delta \mathbf{X}_0' \mathbf{Y}_0 + k\bm{R}\bm{\mu}_0 \right)$, 
$\bm{\beta}^{*}=
\left(\mathbf{X}'\mathbf{X}+
\delta\mathbf{X}_0'\mathbf{X}_0 + k\bm{R} \right)^{-1}
\left(\mathbf{X}'\mathbf{Y} + \delta \mathbf{X}_0' \mathbf{Y}_0 + k\bm{R}\bm{\mu}_0 \right)$,  
\begin{align*}
H_{0}(\delta) &= b + \frac{\delta \left\{ S_0 + k \left(\bm{\mu}_0 -\hat{\bm{\beta}}_0\right)'\mathbf{X}_0'\mathbf{X}_0\left(
\delta\mathbf{X}_0'\mathbf{X}_0 +k \bm{R} \right)^{-1} \bm{R}\left(\bm{\mu}_0 -\hat{\bm{\beta}}_0\right) \right\}}{2}, 
~ {\text and} \\
H(\delta) &= H_{0}(\delta) + 
\frac{  S + \left(\tilde{\bm{\beta}} -\hat{\bm{\beta}}\right)'\mathbf{X}'\mathbf{X}\left( \mathbf{X}'\mathbf{X}+
\delta\mathbf{X}_0'\mathbf{X}_0 +k \bm{R} \right)^{-1}(\delta\mathbf{X}_0'\mathbf{X}_0 +k \bm{R} )
\left(\tilde{\bm{\beta}} -\hat{\bm{\beta}}\right) }{2}.
\end{align*}
\end{result}

It is worth noting that in the case when $\pi_{0}({\bm \beta}, \sigma^2)$ is 
the reference prior, 
the power prior (with fixed $\delta$) for $\delta < p/n_{0}$ is still appropriately 
defined, since after incorporating the full likelihood of the 
current data $D$, the resulting posterior is still proper. 
However, the normalized power prior is only defined with   
 $\mathbfcal{A} = \left(p/n_{0}, 1 \right]$. 
Likewise, when using the empirical Bayes method
to choose $\delta_{\rm EB}$, it must be within the range of 
$\mathbfcal{A}$, while using other 
criterion, for example the DIC, can yield 
different yet more suitable results. This will be illustrated 
via numerical examples below.

\section{Numerical Examples}
\label{sec:simulation}
We conduct two numerical studies to explore how data heterogeneity 
could impact the choice of $\delta$ using different criteria and 
initial priors, and assess its impact on making posterior inference. We primarily illustrate our findings 
with the widely used marginal likelihood criterion, which is also 
equivalent to the use of the posterior mode of $\delta$ with 
the normalized power prior under a uniform initial prior on $\delta$.

\begin{figure}[htbp!]
\begin{center}
\includegraphics[height = 2.8 in, width= 3.3 in]{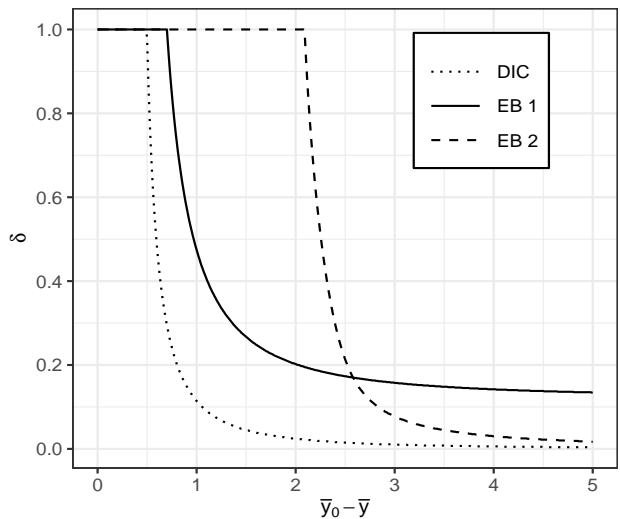}
\end{center}
\vspace{-0.8 cm}
\caption{The selected $\delta$ 
as a function of the sample statistics $\bar{y}_0 -\bar{y}$ in a 
normal population with $n_0 = n = 10$, $s = s_{0} = 0.5$ and 
$\bar{y} = 0$. 
\label{fig:pp-delta}}
\end{figure}

In an intercept--only model 
(normal population without covariates), the criteria (\ref{eq:marginal-lik-LM}) and (\ref{eq:DIC-LM}) can be simplified to  
a function of the sample statistics $\bar{y}_0 - \bar{y}$ when 
the sample variances of $D_0$ and $D$ 
are the same. We set $\bar{y}_0 - \bar{y}$ at 
various levels which creates different degrees of heterogeneity, and 
plot its relationship with the selected $\delta$ in Figure \ref{fig:pp-delta}. 
In this experiment, the sample sizes 
of $D_0$ and $D$ are $n_0 = n = 10$, 
with standard deviations $s = s_{0} = 0.5$ 
respectively, and the 
sample mean of the current data is $\bar{y} = 0$. 
The solid curve (labeled as EB1) 
is the $\delta_{\rm EB}$ 
using reference (initial) prior 
$\pi_{0}(\mu, \sigma^2) \propto 1/\sigma^2$, and 
each point on the dotted curve is the $\delta$ associated with the minimum DIC
with the same reference prior. 
The dashed curve (labeled as EB2) is the 
$\delta_{\rm EB}$ with $\pi_{0}(\mu, \sigma^2) 
= \pi_{0}(\sigma^2) \pi_{0}(\mu  |  \sigma^2)$, 
where $\pi_{0}(\sigma^2) \propto 1/\sigma^2$ and 
$\pi_{0}(\mu  |  \sigma^2) \sim N(0, 10^4\sigma^2)$.
Note that the initial prior used in EB1 is equivalent to 
the use of the prior of the form (\ref{eq:prior-general}) when $p = 1$, with $t = 1$ and $k = b = 0$. The initial prior used in EB2 is also a special case of (\ref{eq:prior-general}) when $p = 1$, with $\mu_0 = b = 0$, $t = 1.5$, $k = 1$, and $R = 10^{-4}$.

The general trend in Figure \ref{fig:pp-delta} 
shows that the optimal $\delta$ will decrease with the increase of 
$\bar{y}_0 - \bar{y}$, which 
reflects the prior--data conflict in an expected way. 
However, the desired $\delta$ should be very close to $0$ when 
the discrepancy between $\bar{y}_0$ and $\bar{y}$ is large. 
This is not the case when using an empirical Bayes with reference prior 
(EB1), since the range of feasible $\delta$ is $(0.1, 1]$, 
which precludes values below $0.1$. In other words, enforcing 
$\delta \in (0.1, 1]$ will possibly result in borrowing more 
information than the optimal choice of $\delta$ under strong 
heterogeneity. This suggests that both the empirical Bayes and the 
normalized power prior should be cautiously used with improper initial priors. 

\begin{figure}[htbp!]
\begin{center}
\includegraphics[height = 2.8 in, width= 3.1 in]{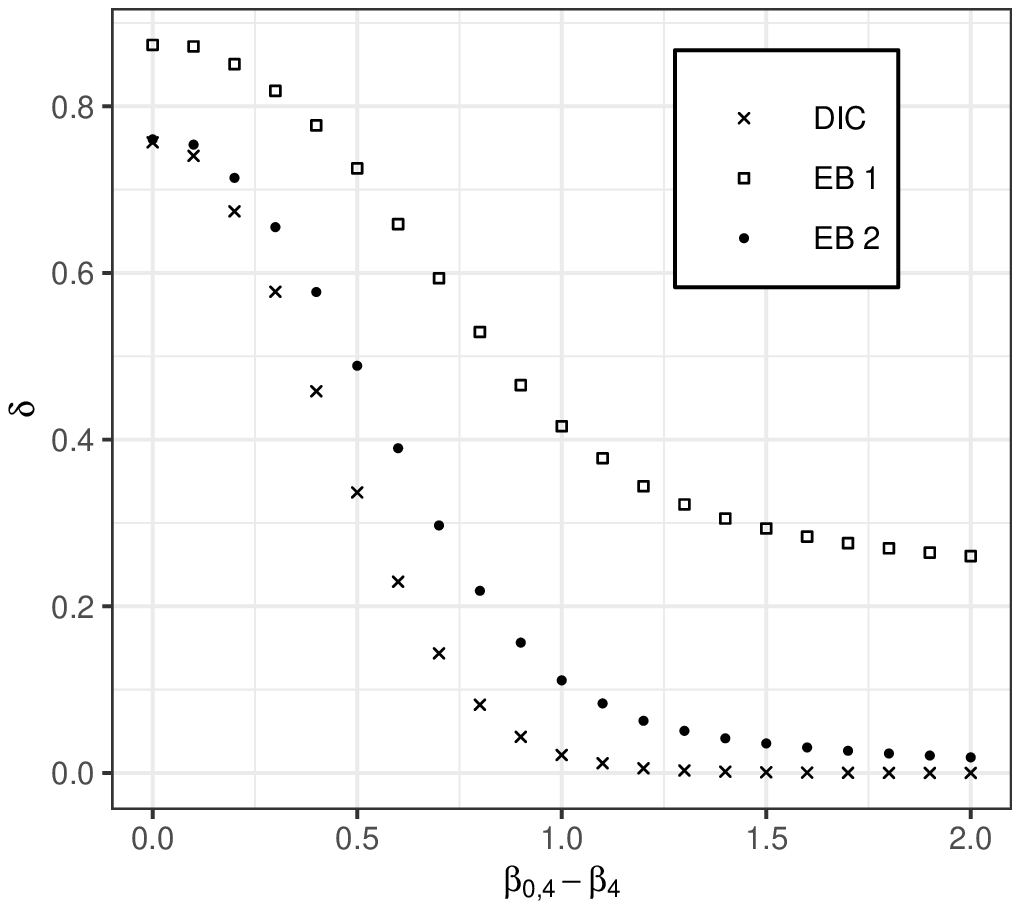} 
\includegraphics[height = 2.8 in, width= 3.1 in]{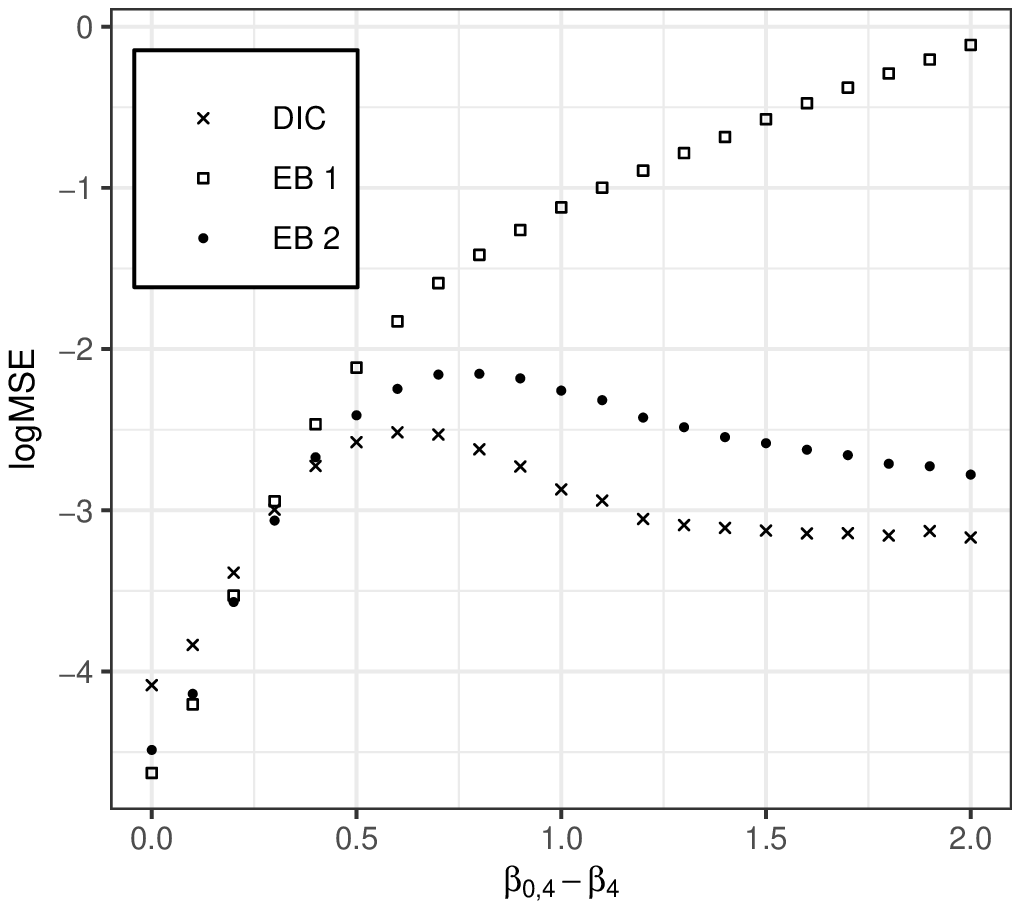} 
\end{center}
\vspace{-0.8 cm}
\caption{
Left: The average values of the selected $\delta$. Right: The corresponding mean squared error for $\hat{\beta}_4$ 
(in log scale), 
with $10^4$ 
data sets simulated from different values of $\beta_{0,4}-\beta_{4}$. 
\label{fig:pp-reg}}
\end{figure}

To assess its impact on the inferential results for  model parameters, in the following experiment, we consider a linear regression model with an intercept and three covariates so the regression parameter is 
$\bm{\beta} = (\beta_1, \beta_2, \beta_3, \beta_4)'$, and the 
variance parameter is $\sigma^2$. 
To generate different levels of 
heterogeneity between the historical and the current data, 
we simulate current data $D$ with 
$\bm{\beta} = (\beta_1, \beta_2, \beta_3, \beta_4)' = (1,1,1,1)'$ and 
simulate historical data $D_0$ with 
$\bm{\beta} = (1, 1, 1, \beta_{04})'$,
where $\beta_{04}$ takes a grid of the values between $1$ and $3$. 
We generate $10^4$ data sets for each scenario, with sample sizes 
$n = n_0 = 20$ and the covariates are generated from the uniform 
distribution on $(0, 1)$. 
For each dataset, we first choose the optimal $\delta$ with criteria 
similar to those in Figure \ref{fig:pp-delta}, 
and display the results of the average $\delta$ over the $10^4$ samples 
in Figure \ref{fig:pp-reg} (left). For each data set, 
we also calculate the posterior mean of ${\bf \beta}_{4}$ 
with the power priors using the corresponding $\delta$, 
and report the logarithm of the mean squared error (logMSE) in Figure \ref{fig:pp-reg} (right). 
Likewise, the empirical Bayes with the reference prior of 
the form $\pi_{0}({\bm \beta}, \sigma^2) 
\propto 1/\sigma^2$ is denoted as EB1, and the empirical Bayes
with the prior of the form 
$\pi_{0}({\bm \beta}, \sigma^2) \propto \pi_{0}(\sigma^2) \pi_{0}
({\bm \beta}  |  \sigma^2)$, 
where $\pi_{0}(\sigma^2) \propto 1/\sigma^2$ and 
$\pi_{0}({\bm \beta}  |  \sigma^2) 
\sim {\rm N}_{4}(\bm{0}_{4}, 10^4 \sigma^2 I_4)$, is denoted as 
EB2. 
The DIC is obtained by using the reference prior described above 
as the initial prior, but selecting the optimal $\delta$ via the DIC. 
Setting $\delta$ as random and using the normalized power prior 
(with $\pi_{0}(\delta) \sim {\rm unif}(0,1)$) 
with (initial) reference prior provides similar results to the EB1, so it is not shown in the plots. 

Similar to Figure \ref{fig:pp-delta}, we observe from  Figure \ref{fig:pp-reg} (left) that on average, 
the borrowing strength will decrease  
when the level of heterogeneity between $D_0$ and $D$ increases, but 
the lower limit is $p/n_0 = 0.2$ when a reference prior is used 
with the empirical Bayes (this is also the lower bound of $\delta$ if using the normalized power prior). 
Therefore, when estimating the parameter $\beta_4$, at 
least $20\%$ of the information will be borrowed on average 
from historical data 
regardless of the strength of heterogeneity. 
As a result, the logMSE increases monotonically with 
$\beta_{0,4}-\beta_{4}$. 
On the other hand, the lower limit of $\delta$ in  the other two 
approaches is $0$, so eventually their logMSE can be maintained to 
the same level as no borrowing. 

Overall we can conclude that, when using either 
the empirical Bayes with a fixed $\delta$, or the normalized power prior 
in a normal linear model, an improper initial prior may prevent the 
selection of an optimal $\delta$. The primary example is the 
reference prior, which is commonly used in  Bayesian analysis with the
full likelihood. 
On the other hand, under these scenarios, 
the DIC and other information criteria could be better alternatives.



\section{Concluding Remarks}
\label{sec:conc}
In this study, we have discussed a critical yet largely ignored key point for the use of the power priors or its modified form, the normalized power prior,  
to borrow information from historical data in conducting a new study. 
By establishing general results 
regarding the propriety of a powered posterior 
(i.e., using a fractional likelihood with power $\delta$) 
under various initial priors, 
we showed that the lower limit for $\mathbfcal{A}$, the feasible set of $\delta$, is 
not necessarily $0$ with an improper initial prior, 
even if an initial  
prior can yield a proper posterior in conventional Bayesian 
inferences (i.e., $\delta = 1$). 
We thus advocated the use of 
a more rigorous formulation for the normalized power prior, 
as well as for the formula of the marginal likelihood. 
These formulations are provided 
in (\ref{eq:marginal-lik}) and (\ref{eq:npp-redef}), which account for 
the aforementioned restriction on the power parameter. 

What is the influence of this result on Bayesian inference? We showed in a normal linear model, with the reference prior as the initial prior, 
the lower limit of feasible 
$\delta$ is $p/n_{0}$ when using the normalized power prior or the 
marginal likelihood criterion. 
Such impact could have a strong bearing on parameter estimates especially when 
the sample size of the historical data is small or moderate 
with some covariates, whereas the heterogeneity between the historical and the current data is strong.
Therefore, an ideal level of borrowing 
may fall below the lower limit $p/n_{0}$. 
This might be encountered especially in the case 
when one splits the whole historical dataset into multiple small ones and borrows each of them individually \citep{Banbeta19}. 
In these cases, we should avoid using either the 
marginal likelihood criterion or the normalized power prior and choose other criteria instead. 
In reality, it is also suggested to use multiple information criteria when possible, while another option is to use a (vague) proper initial prior. 

More generally, sampling from the posterior based on a fractional likelihood 
 is not only used for informative prior elicitation, but also a 
technique widely used in Bayesian computation for more generic problems. 
For instance, \cite{Friel08} calculated the normalizing 
constant based on ideas of thermodynamic integration or path sampling \citep{GelmanMeng98} 
with the identity
\begin{align*}\label{eq:friel}
\log z 
=\int_{\mathcal{A}}
E_{\pi(\boldsymbol{\theta}  \mid  D, \delta^* )} 
\{ \log L( \boldsymbol{\theta} \mid D ) \}  d {\delta^*},
\end{align*}
where the feasible set $\mathcal{A}$ is similarly defined as in (\ref{eq:marginal-lik}). 
 We use the same notation $z$ as in 
\cite{GelmanMeng98} to denote the normalizing 
constant in a general Bayesian model, and 
$L( \boldsymbol{\theta} | D )$ is the likelihood based on the data $D$. 
Note that under this general setting we do not  consider the historical data $D_0$, 
so essentially $D$ takes the role of $D_0$ in 
(\ref{eq:pp}), and now (\ref{eq:pp}) becomes a powered posterior. Then 
$\pi(\boldsymbol{\theta} |  D, \delta^* ) 
\propto \pi({\bm \theta}) L({\bm \theta} |D)^{\delta^*}$, where $\pi(\boldsymbol{\theta})$ denotes an arbitrary prior for $\boldsymbol{\theta}$
such that 
$z = \int_{\bm \Theta} 
\pi(\boldsymbol{\theta}) 
L( \boldsymbol{\theta} | D ) d {\bm \theta}
$. The integrand is 
approximated by  
sampling from a sequence of the posterior densities 
based on different powered likelihoods 
with power ${\bm t} = \{ t_i\}_{i=1}^{s} \in \mathcal{A}$, where 
$0 \leq t_1 < \ldots < t_s \leq 1$, and the 
integral is approximated by the trapezoidal rule. 
Often $\pi({\bm \theta})$ is assumed to be proper such that 
the sequence of the corresponding powered 
posteriors is believed to be proper, 
while Theorem \ref{thm:proper-prior} provided the evidence 
towards this belief. 
When an improper prior is used, our result 
indicated that the starting point $t_1$ is not 
necessarily very close to $0$. If an improper prior that 
can yield a proper posterior based on the powered 
likelihood with $t_1 > 0$, Theorem  \ref{rmk:improper-prior-proper-region} demonstrated that 
sampling from all the subsequent powered posteriors 
with ${\bm t}$ is valid. 

\section*{Acknowledgement}
The authors would like to express our deep appreciation for the Associate Editor and the anonymous reviewer for their 
comments and suggestions, which lead to a much improved article.

\section*{Appendix: Additional Examples and Proofs of Theorems}
\label{sec:appendix}
\renewcommand{\theequation}{A.\arabic{equation}}

\noindent{\bf An example to illustrate the 
differences between (\ref{eq:jpp}) and 
(\ref{eq:npp-redef})}
For independent Bernoulli trials with $y_0$ successes out of the $n_0$ trials in $D_0$, suppose the probability of success is $\theta$. 
Then the historical likelihood based on the product of independent Bernoulli densities is 
$L(\theta |D_0)= \theta^{y_0} (1-\theta)^{n_0-y_0}$. 
Assuming that $\pi_{0}(\theta) \sim \text{Beta}(a_1, a_2)$, with $a_1 > 0$, $a_2 > 0$, and $\pi_0(\delta)$ is a proper prior 
for $\delta$, the joint power prior using (\ref{eq:jpp}) is of the form 
$$
\pi_{J}(\theta, \delta \mid D_0) \propto \pi_{0}(\delta) \theta^{\delta y_0+a_1-1}(1-\theta)^{\delta(n_0-y_0)+a_2-1}, 
$$
where $\pi_{J}(\theta, \delta|D_0)$ 
stands for a joint power prior. 
If we use the normalized power prior 
(\ref{eq:npp-redef}), the prior is of the form 
$$
\pi_{N}(\theta, \delta \mid D_0) \propto \pi_{0}(\delta) 
\frac{\theta^{\delta y_0+a_1-1}(1-\theta)^{\delta(n_0-y_0)+a_2-1}}
{B(\delta y_{0}+a_1,\delta(n_{0}-y_{0})+a_2)}.
$$
These two priors differ by a multiplicative factor $B(\delta y_{0}+a_1,\delta(n_{0}-y_{0})+a_2)$ which is not a constant. 
Furthermore, if we use the likelihood based on the sufficient statistics 
$y_0$, which follows a binomial distribution, 
the joint power prior denoted as 
$\pi^{*}_{J}(\theta, \delta |  D_0)$ is of
the form 
$$\pi^{*}_{J}(\theta, \delta \mid  D_0) \propto 
\binom{n_0}{y_0}^{\delta} \pi_{J}(\theta, \delta \mid D_0).$$
This is clearly different from $\pi_{J}(\theta, \delta |  D_0)$, which indicates a violation of the likelihood principle \citep{Birnbaum62,Duan06a,Neuenschwander09}. 
The normalized power prior remains unchanged since the extra term cancels in the numerator and denominator.

\medskip

\noindent{\bf Proof of Theorem \ref{thm:proper-prior}}
\noindent 
Assume regularity conditions hold 
including $L({\bm \theta} | D_{0})$ is 
non--negative and finite, $\pi_{0}({\bm \theta}) \geq 0$ and is proper, and 
$P(L({\bm \theta} | D_{0}) > 0)$ is positive. 
Since the function $g(x) = x^{\delta}$ is concave $(0 < \delta < 1)$, from Jensen's inequality we have 
$$
C(\delta) = E_{\pi_{0}({\bm \theta})}\left\{L({\bm \theta} \mid D_{0})^{\delta}\right\}
\leq \left[E_{\pi_{0}({\bm \theta})} \{L({\bm \theta} \mid D_{0})\} \right]^{\delta}, 
$$
which indicates 
\begin{equation}
\label{eq:Jensen-proper}
\int_{\bm \Theta} \pi_{0}({\bm \theta}) L({\bm \theta} \mid D_{0})^{\delta} 
d{\bm \theta} \leq 
\left\{\int_{\bm \Theta} \pi_{0}({\bm \theta}) L({\bm \theta} \mid D_{0})
d{\bm \theta} \right\}^{\delta}.
\end{equation}
Since $\pi_{0}({\bm \theta})$ is proper, $C(0) = 1$. Also, 
the posterior based on the historical data is proper almost surely, i.e. 
$\int_{\bm \Theta} \pi_{0}({\bm \theta}) L({\bm \theta} | D_{0}) d{\bm \theta} 
< \infty$. Therefore $C(\delta)$ is finite when $\delta \in [0, 1]$. 
Note that this proof is used in the proof of Theorem 1 of 
\cite{Carvalho20}, however, its generalization given in  
the \citet[Remark 3]{Carvalho20} may not hold, for the reason stated below. 
\medskip 

\noindent{\bf Some explanations on 
the Remark in Section \ref{sec:meth}}
\noindent 
We first show that the inequality in (\ref{eq:Jensen-proper}) 
in the proof of Theorem \ref{thm:proper-prior} is no 
longer valid if the density $\pi_{0}({\bm \theta})$ is not a 
valid (normalized) probability density function. 
Suppose $\pi_{0}({\bm \theta}) = c \pi^{*}({\bm \theta})$, 
where $\int_{\bm \Theta} \pi^{*}({\bm \theta}) d {\bm \theta} = 1$, 
and $c$ is a positive number. 
The Jensen's inequality indicates that 
$$
E_{\pi^{*}({\bm \theta})}\left\{L({\bm \theta} \mid D_{0})^{\delta}\right\}
\leq \left[E_{\pi^{*}({\bm \theta})} \{L({\bm \theta} \mid D_{0})\} \right]^{\delta}, 
$$
which implies 
\begin{align*}
\int_{\bm \Theta} 
\pi_{0}({\bm \theta})
L({\bm \theta} \mid D_{0})^{\delta} 
d{\bm \theta} &\leq c^{1-\delta}
\left\{\int_{\bm \Theta} \pi_{0}({\bm \theta})
 L({\bm \theta} \mid D_{0})
d{\bm \theta} \right\}^{\delta}.
\end{align*}
This does not satisfy the inequality in (\ref{eq:Jensen-proper}). 
Moreover, when $c$ is not finite, Jensen's inequality 
fails to provide a finite upper bound.
Therefore to utilize the Jensen's inequality, one has to consider 
the normalizing constant, and therefore the result in
(\ref{eq:Jensen-proper}) cannot be generalized to an improper prior $\pi_{0}({\bm \theta})$. 
A primary example to illustrate that the powered posterior based on $D_{0}$ 
can be either proper or improper with an improper initial prior is given in the last paragraph of Section \ref{sec:linear-model} in a normal linear model.



\medskip

\noindent{\bf Proof of Theorem \ref{rmk:improper-prior-proper-region}} 
\noindent Let $\pi({\bm \theta} | D_0, \delta^{*}) = 
\pi_{0}({\bm \theta}) 
L({\bm \theta}  |  D_0)^{\delta^*}C(\delta^*)^{-1}$, 
which is assumed to be proper, and 
$C(\delta^*) = \int_{\bm{\Theta}} \pi_{0}({\bm \theta}) 
L({\bm \theta}  |  D_0)^{\delta^*} < \infty$. 
Now set $\delta_{d} = \delta - \delta^{*}$, 
where $\delta_{d} \in (0, 1)$. We have
\begin{align*}
\int_{\bm \Theta} 
\pi_{0}({\bm \theta})
L({\bm \theta}  \mid  D_0)^{\delta}  d {\bm \theta} 
&= \int_{\bm \Theta} \pi_{0}({\bm \theta})
L({\bm \theta}  \mid  D_0)^{\delta^*} 
L({\bm \theta}  \mid  D_0)^{\delta_{d}} d {\bm \theta} \nonumber \\
&= C(\delta^*) E_{\pi({\bm \theta} | D_0, \delta^{*})}
\left\{ L({\bm \theta}  \mid D_0)^{\delta_{d}} \right\}.
\end{align*}
Since ${\pi({\bm \theta} | D_0, \delta^{*})}$ is 
the probability density function and $\delta_{d} \in (0, 1)$, from 
Theorem \ref{thm:proper-prior}, 
$E_{\pi({\bm \theta} | D_0, \delta^{*})}
\left\{ L({\bm \theta}  |  D_0)^{\delta_{d}} \right\}$ 
is finite. Therefore 
$\int_{\bm \Theta} 
\pi_{0}({\bm \theta})
L({\bm \theta}  |  D_0)^{\delta}  d {\bm \theta}$ is also finite, which completed the proof. 

\medskip

\noindent{\bf Proof of Result \ref{thm:normal-model}} 
With likelihood of the form
$$L(\bm{\beta},\sigma^2 \mid D_{0})\propto 
 \frac{1}{(2\pi \sigma^2)^{\frac{n_{0}}{2}}}\exp\left[- \frac{1}{2 \sigma^2} 
 \left\{S_{0} + (\bm{\beta}-\hat{\bm{\beta}}_{0})'\mathbf{X}_{0}'
 \mathbf{X}_{0}(\bm{\beta}-
 \hat{\bm{\beta}}_{0}) \right\} \right],
$$
where $\hat{\bm{\beta}}_0=(\mathbf{X}_{0}'\mathbf{X}_{0})^{-1}
    \mathbf{X}_{0}'\mathbf{Y}_{0}$
   and 
${S}_0 = (\mathbf{Y}_0-\mathbf{X}_{0} \hat{\bm{\beta}}_0)'(\mathbf{Y}_0-\mathbf{X}_{0} \hat{\bm{\beta}}_0)$, 
we have
\begin{align*}
\label{eq:C-delta}
C(\delta) &= \int_{0}^{\infty} \int_{\mathcal{R}^p} \pi_{0}(\bm{\beta},\sigma^2)
L(\bm{\beta},\sigma^2 | D_0)^{\delta}d\bm{\beta}d\sigma^2 \nonumber \\
&= 
\int_{0}^{\infty} \int_{\mathcal{R}^p} 
\frac{(2\pi)^{-\frac{n_{0}\delta}{2}}}{(\sigma^2)^{t+\frac{n_0 \delta}{2}}}
\exp\left\{-\frac{
(\bm{\beta} -\tilde{\bm{\beta}})' (\delta
\mathbf{X}_0'\mathbf{X}_{0} +k \bm{R})(\bm{\beta} -\tilde{\bm{\beta}}) + 
2H_{0}(\delta)}{2\sigma^2} \right\} d\bm{\beta}d\sigma^2 
\nonumber \\ &= 
(2\pi)^{-\frac{n_{0}\delta-p}{2}} \Gamma(\nu_0) |\delta \mathbf{X}_0'\mathbf{X}_{0} + k \bm{R}|^{-\frac{1}{2}} 
H_{0}(\delta)^{-\nu_0},
\end{align*}
where $\nu_0$, 
$\tilde{\bm{\beta}}$, and 
$H_{0}(\delta)$ are defined as per Result 
\ref{thm:normal-mlcdic}.
The second line follows from completing the squares of the form 
\begin{align*}
&(\bm{\beta} -{\bm{\mu}}_{0})'  k\bm{R} (\bm{\beta}-{\bm{\mu}}_{0}) + 
(\bm{\beta} -\hat{\bm{\beta}}_{0})' \delta
\mathbf{X}_0'\mathbf{X}_{0} (\bm{\beta} -\hat{\bm{\beta}}_{0}) \\
=&
(\bm{\beta} - \tilde{\bm{\beta}})' (\delta
\mathbf{X}_0'\mathbf{X}_{0} + k\bm{R})(\bm{\beta} - \tilde{\bm{\beta}})
+ \delta k \left(\bm{\mu}_0 -\hat{\bm{\beta}}_0\right)'\mathbf{X}_0'\mathbf{X}_0\left(
\delta\mathbf{X}_0'\mathbf{X}_0 +k \bm{R} \right)^{-1} \bm{R}\left(\bm{\mu}_0 -\hat{\bm{\beta}}_0\right), 
\end{align*}
and the last line follows from using the multivariate normal-inverse-gamma integral. 
Clearly, $C(\delta)$ is finite when 
$\nu_0 > 0$. Thus we have 
$\mathcal{A} = \left\{ \delta  \mid \delta > 
(2-2t+p)/n_0 \right\}$. 
When $t = 1$ (as in the reference prior), clearly $\delta$ is defined only when $\delta > {p}/{n_0}$. 
When $t \geq 1 + p/2$, $\delta$ is defined on 
$(0, 1]$. 

\medskip

\noindent{\bf Proofs of Result 
\ref{thm:normal-mlcdic}}
The normalized power prior $\pi({\bm \beta}, \sigma^2, \delta  |  D_0) $ is proportional to 
\begin{align*}
\left(\frac{1}{\sigma^2}\right)^{\frac{n_0 \delta}{2} + t} 
\frac{\pi_{0}(\delta) H_{0}(\delta)^{\nu_0}
}{\Gamma(\nu_0) |\delta\mathbf{X}_0'\mathbf{X}_0 +k \bm{R}|^{-\frac{1}{2}}}
\exp\left\{-\frac{
(\bm{\beta} -\tilde{\bm{\beta}})' (\delta
\mathbf{X}_0'\mathbf{X}_{0} +k \bm{R})(\bm{\beta} -\tilde{\bm{\beta}}) +
2H_{0}(\delta)}{2\sigma^2} \right\}.
\end{align*}
Multiplying by the likelihood of the current data $L({\bm \beta}, \sigma^2  | D)$, and by a similar 
argument, the full posterior 
$\pi({\bm \beta}, \sigma^2 , \delta  |  D, D_{0})$ is proportional to 
\begin{equation}
\left(\frac{1}{\sigma^2}\right)^{\frac{n+n_0 \delta}{2} + t} 
\frac{\pi_{0}(\delta) H_{0}(\delta)^{\nu_0} 
}{\Gamma(\nu_0 ) |\delta\mathbf{X}_0'\mathbf{X}_0 +k \bm{R}|^{-\frac{1}{2}}}
\exp\left\{-\frac{
(\bm{\beta} -{\bm{\beta}}^{*})' (\mathbf{X}'\mathbf{X} + \delta
\mathbf{X}_0'\mathbf{X}_{0} +k \bm{R})(\bm{\beta} -{\bm{\beta}}^{*}) + 
2H(\delta)}{2\sigma^2} \right\},
\label{eq:full-posterior}
\end{equation}
where $H(\delta)$ and $\bm{\beta}^{*}$ are defined in Result \ref{thm:normal-mlcdic}. 
To get the marginal posterior $\pi(\delta  |  D_0, D)$ 
(where $\delta \in \mathcal{A}$), 
we integrate $({\bm \beta}, \sigma^2)$ out from (\ref{eq:full-posterior}), 
which is of the form
\begin{equation*}
\pi(\delta \mid D_0, D) \propto \frac{\Gamma(\nu) |\delta
\mathbf{X}_0'\mathbf{X}_{0} +k \bm{R}|^{\frac{1}{2}} H_{0}(\delta)^{\nu_0} \pi_{0}(\delta)}
{\Gamma(\nu_{0}) | \mathbf{X}'\mathbf{X} + \delta
\mathbf{X}_0'\mathbf{X}_{0} +k \bm{R}|^{\frac{1}{2}} H(\delta)^{\nu}}, 
\end{equation*}
where $\nu$ is defined in Result \ref{thm:normal-mlcdic}.
Setting $\pi_0(\delta) = 1$ 
we can easily derive the marginal likelihood 
$m(\delta | D_0, D)$. 
\\
The DIC calculation: 
For given $\delta \in {\mathcal A}$, the conditional posterior of $({\bm \beta}, \sigma^2  | D, D_0, \delta)$ is 
\begin{equation*}
\pi({\bm \beta}, \sigma^2  \mid  D, D_0, \delta) \propto 
\left(\frac{1}{\sigma^2}\right)^{\frac{n+n_0 \delta}{2} + t} 
\exp\left\{-\frac{
(\bm{\beta} -{\bm{\beta}}^{*})' (\mathbf{X}'\mathbf{X} + \delta
\mathbf{X}_0'\mathbf{X}_{0} +k \bm{R})(\bm{\beta} -{\bm{\beta}}^{*}) + 
2H(\delta)}{2\sigma^2} \right\} ,
\end{equation*}
which is the normal-inverse-gamma kernel with ${\rm  N}_{p} \mbox{-} \Gamma^{-1}({\bm \beta}^{*}, 
\mathbf{X}'\mathbf{X} + \delta \mathbf{X}_0'\mathbf{X}_{0} +k \bm{R}, \nu, H(\delta))$. Therefore 
\begin{equation}
E({\bm \beta} \mid D, D_0, \delta) = {\bm \beta}^{*}, \;\;\;  E(\sigma^2  \mid  D, D_0, \delta) = \frac{H(\delta)}{\nu - 1}. 
\label{eq:expected-theta}
\end{equation}
Recall the definition of deviance for parameter ${\bm \theta}$, defined as 
$
{\rm Dev}_{L({\bm \theta}  |  D)}({\bm \theta}) = -2 \log L({\bm \theta}  |  D). $
Hereafter we use the subscript 
$L({\bm \theta} | D)$ to clarify that the 
deviance is for the current data model. 
Then  
\begin{align*}
&{\rm Dev}_{L({\bm \beta}, \sigma^2 |D)} \{ E({\bm \beta}, \sigma^2  \mid D, D_0, \delta) \} \\
= & ~
n\left\{\log(2\pi) + \log H(\delta) - \log(\nu-1) \right\}
+ \frac{\nu-1}{H(\delta)}
\left\{ ({\bm \beta}^{*} - \hat{\bm \beta})' 
\mathbf{X}'\mathbf{X} ({\bm \beta}^{*} - \hat{\bm \beta})
+ S \right\}.
\end{align*}
Also
\begin{align}
&~ ~ ~ ~E\left\{{\rm Dev}_{L({\bm \beta}, \sigma^2 |D)}({\bm \beta}, \sigma^2)  \mid  D, D_0, \delta \right\} 
\nonumber \\
&=  -2E_{\pi({\bm \beta}, \sigma^2  |  D, D_0, \delta)} \left\{ \log L({\bm \beta}, \sigma^2  \mid  D) \right\} \nonumber \\
&=  n\log(2\pi) + E_{\pi({\bm \beta}, \sigma^2  |  D, D_0, \delta)} \left\{ n \log(\sigma^2) + \frac{S}{\sigma^2}
+ \frac{(\bm{\beta} -\hat{\bm{\beta}})' \mathbf{X}'\mathbf{X}(\bm{\beta} - \hat{\bm{\beta}})}{\sigma^2} \right\} \nonumber \\
&=  n\left\{ \log(2\pi) + \log H(\delta) - \psi(\nu)  \right\} +  \frac{\nu}{H(\delta)}
\left\{ ({\bm \beta}^{*} - \hat{\bm \beta})' 
\mathbf{X}'\mathbf{X} ({\bm \beta}^{*} - \hat{\bm \beta})
+ S \right\}
\nonumber \\
&~ ~ ~ ~ +
{\rm tr}(\mathbf{X}'\mathbf{X} (\mathbf{X}'\mathbf{X} + \delta \mathbf{X}_0'\mathbf{X}_{0} +k \bm{R})^{-1}),
\label{eq:E-Deviance}
\end{align}
where $\psi(\cdot)$ is the digamma function. 
The  first two terms in 
(\ref{eq:E-Deviance}) other than the $n\log(2\pi)$ can be derived using 
the fact that the marginal 
posterior $\pi(\sigma^2  |D_0, D)$ follows an 
inverse gamma distribution with shape $\nu$ and scale 
$H(\delta)$. For the last term 
$E_{\pi({\bm \beta}, \sigma^2  |  D, D_0, \delta)} \big\{ 
(\bm{\beta} -\hat{\bm{\beta}})' \mathbf{X}'\mathbf{X}(\bm{\beta} - \hat{\bm{\beta}})/{\sigma^2} \big\}$, 
 the integrand is 
\begin{equation}
\frac{|{\bm \Lambda}|^{\frac{1}{2}} H(\delta)^{\nu}}
{(2\pi)^{\frac{p}{2}} \Gamma(\nu) }
\left( \frac{1}{\sigma^2}\right)^{\nu+\frac{p}{2}+2}
\exp\left\{-\frac{(\bm{\beta} -{\bm{\beta}^*})' {\bm \Lambda}
(\bm{\beta} -{\bm{\beta}^*}) + 2H(\delta)}{2\sigma^2}\right\}
(\bm{\beta} -\hat{\bm{\beta}})' \mathbf{X}'\mathbf{X}(\bm{\beta} - \hat{\bm{\beta}}),
\label{eq:DIC-integrand1}
\end{equation}
where ${\bm \Lambda} = \mathbf{X}'\mathbf{X} + \delta \mathbf{X}_0'\mathbf{X}_{0} +k \bm{R}$. We first integrate 
$\sigma^2$ out from (\ref{eq:DIC-integrand1}) which results in \begin{equation}
\frac{|{\bm \Lambda}|^{\frac{1}{2}} \Gamma(\nu+\frac{p}{2}+1)}
{(2\pi)^{\frac{p}{2}} H(\delta)^{\frac{p}{2}+1}\Gamma(\nu) }
\left\{
1 + \frac{(\bm{\beta} -{\bm{\beta}^*})' {\bm \Lambda}
(\bm{\beta} -{\bm{\beta}^*})}{2 H(\delta)}
\right\}^{-(\nu+\frac{p}{2}+1)}
(\bm{\beta} -\hat{\bm{\beta}})' \mathbf{X}'\mathbf{X}(\bm{\beta} - \hat{\bm{\beta}}).
\label{eq:DIC-integrand2}
\end{equation}
Let ${\bm \Sigma} = \frac{H(\delta)}{\nu+1}{\bm \Lambda}^{-1}$ and $\nu^{*}= 2\nu+2$. Equation 
(\ref{eq:DIC-integrand2}) can be expressed as 
$$
\frac{|{\bm \Sigma}|^{-\frac{1}{2}} \Gamma\left(
\frac{\nu^*+p}{2}\right)}{\Gamma(\nu^{*}/2) 
(\pi \nu^{*})^{\frac{p}{2}}}
\left\{
1 + \frac{1}{\nu^{*}}
(\bm{\beta} -{\bm{\beta}^*})' {\bm \Sigma}^{-1}
(\bm{\beta} -{\bm{\beta}^*})
\right\}^{-\frac{\nu^* + p}{2}}\frac{\nu}{H(\delta)}
(\bm{\beta} -\hat{\bm{\beta}})' \mathbf{X}'\mathbf{X}(\bm{\beta} - \hat{\bm{\beta}}),
$$
which is $\nu H(\delta)^{-1}
(\bm{\beta} -\hat{\bm{\beta}})' \mathbf{X}'\mathbf{X}(\bm{\beta} - \hat{\bm{\beta}})$ 
multiplies a multivariate Student-$t$ density 
with location parameter ${\bm \beta}^{*}$, 
shape matrix ${\bm \Sigma}$ and 
the degree of freedom $\nu^{*}$. 
Applying the expectation of a quadratic form we 
get (\ref{eq:E-Deviance}).

Combining (\ref{eq:expected-theta}) and 
(\ref{eq:E-Deviance}) we can easily derive an analytical form 
of the DIC. 
The effective number of parameters \citep{Spiegelhalter02} 
in our model is denoted by 
\begin{align*}
p_{D}(\delta | D_0, D) &= 
E\left\{ {\rm Dev}_{L({\bm \beta}, \sigma^2 |D)}({\bm \beta}, \sigma^2)  \mid  D, D_0, \delta \right\} - 
{\rm Dev}_{L({\bm \beta}, \sigma^2 |D)}\{ E({\bm \beta}, \sigma^2  \mid  D, D_0, \delta) \} 
\\
&= 
n(\log(\nu-1) - \psi(\nu)) +  
\frac{1}{H(\delta)}
\left\{(\bm{\beta}^{*} -\hat{\bm{\beta}})' \mathbf{X}'\mathbf{X}(\bm{\beta}^{*} - \hat{\bm{\beta}}) + S
\right\}
\\& ~ ~ ~ ~ + 
{\rm tr}(\mathbf{X}'\mathbf{X} (\mathbf{X}'\mathbf{X} + \delta \mathbf{X}_0'\mathbf{X}_{0} +k \bm{R})^{-1}). 
\label{eq:p_D}
\end{align*}
Now the DIC for a model with specific $\delta$, 
up to a constant, is given by 
\begin{align*}
{\rm DIC}(\delta | D_0, D) &= 
{\rm Dev}_{L({\bm \beta}, \sigma^2 |D)} \{ E({\bm \beta}, \sigma^2  \mid  D, D_0, \delta) \} + 
2 p_{D}(\delta | D_0, D) 
\\ & =
n\left\{\log(\nu-1)+ \log H(\delta) -2\psi(\nu)\right\} + 
\frac{\nu+1}{H(\delta)} \left\{
(\bm{\beta}^{*} -\hat{\bm{\beta}})' \mathbf{X}'\mathbf{X}(\bm{\beta}^{*} - \hat{\bm{\beta}}) 
+ S\right\}
\\& ~ ~ ~ ~ + 
2{\rm tr}(\mathbf{X}'\mathbf{X} (\mathbf{X}'\mathbf{X} + \delta \mathbf{X}_0'\mathbf{X}_{0} +k \bm{R})^{-1}). 
\end{align*}
This completed the derivation of Result \ref{thm:normal-mlcdic}.

\bibliographystyle{chicago}

\bibliography{Bibliography-MM-MC}
\end{document}